\magnification=\magstep1 
\font\bigbfont=cmbx10 scaled\magstep1
\font\bigifont=cmti10 scaled\magstep1
\font\bigrfont=cmr10 scaled\magstep1

\vsize = 23.5 truecm
\hsize = 15.5 truecm
\hoffset = .2truein
\baselineskip = 14 truept
\overfullrule = 0pt
\parskip = 3 truept
\def\bfR{{\bf R}}
\def\bfr{{\bf r}}
\def\bfk{{\bf k}}

\def\perm{\mathop {\rm Perm}}
\def\gamcc{\Gamma_{cc}}
\def\bcc{B_{cc}}

\def\tbul#1{{\mathop {#1}\limits^\bullet}}
\def\rij{r_{12}}
\centerline{\bigbfont BOSE-EINSTEIN CONDENSATION}
\vskip 8 truept
\centerline{\bigbfont AND THE LAMBDA TRANSITION IN LIQUID HELIUM}
\vskip 23 truept
\centerline{\bigifont T. Lindenau and M. L. Ristig}
\vskip 9 truept
\centerline{\bigrfont Institut f\"ur Theoretische Physik} 
\vskip 3 truept
\centerline{\bigrfont Universit\"at zu K\"oln, D-50937 K\"oln, Germany}
\vskip 16 truept
\centerline{\bigifont J. W. Clark}
\vskip 9 truept
\centerline{\bigrfont McDonnell Center for the Space Sciences} 
\vskip 3 truept
\centerline{\bigrfont and Department of Physics, Washington University}
\vskip 3 truept
\centerline{\bigrfont St. Louis, Missouri 63130, USA} 
\vskip 16 truept
\centerline{\bigifont K. A. Gernoth}
\vskip 9 truept
\centerline{\bigrfont Department of Physics, UMIST}
\vskip 3 truept
\centerline{\bigrfont P.O. Box 88, Manchester M60 1QD, United Kingdom} 
\vskip 1.2 truecm
\noindent
{\bf Abstract:} $\qquad$
Integrating seminal ideas of London, Feynman, Uhlenbeck, Bloch, 
Bardeen, and other illustrious antecessors, this paper continues the 
development of an {\it ab initio} theory of the $\lambda$ transition in
liquid $^4$He.  The theory is based upon variational determination of 
a correlated density matrix suitable for description of both normal
and superfluid phases, within an approach that extends to finite 
temperatures the very successful correlated wave-functions theory of 
the ground state and elementary excitations at zero temperature.  We 
present the results of a full optimization of a correlated trial 
form for the density matrix that includes the effects both of 
temperature-dependent dynamical correlations and of statistical 
correlations corresponding to thermal phonon/roton and quasiparticle/hole 
excitations -- all at the level of two-point descriptors.  The
optimization process involves constrained functional minimization of 
the associated free energy through solution of a set of Euler-Lagrange 
equations, consisting of a generalized paired-phonon equation for 
the structure function, an analogous equation for the Fourier 
transform of the statistical exchange function, and a Feynman 
equation for the dispersion law of the collective excitations.  
Violation of particle-hole exchange symmetry emerges as an important 
aspect of the transition, along with broken gauge symmetry.  In 
conjunction with a semi-phenomenological study in which 
renormalized masses are introduced for quasiparticle/hole and 
collective excitations, the results suggest that a quantitative 
description of the $\lambda$ transition and associated thermodynamic
quantities can be achieved once the trial density matrix is modified 
-- notably through the addition of three-point descriptors -- to 
include backflow effects and allow for {\it ab initio} treatment 
of important variations in effective masses.
\vfill\eject

\centerline{\bf 1.  INTRODUCTION}
\vskip 12 truept
The remarkable coherent collective property of superfluidity, observable 
in liquid $^4$He below the famous $\lambda$ transition, was discovered by
Kapitza [1] and Allen and Misener [2] in 1938.  Almost immediately, 
Fritz London [3,4] proposed that the transition to the superfluid
phase (Helium II) reflects a process that is fundamentally analogous 
to Bose-Einstein condensation of an ideal Bose gas:  
{\par\narrower\noindent
{\sl
I recently realized that ... some support could be given to the
idea that the peculiar phase transition (``$\lambda$-point''),
that liquid helium undergoes at 2.19~K, very probably has to be
regarded as the condensation phenomenon of the Bose-Einstein
statistics, distorted, of course, by the presence of molecular forces
and by the fact that it manifests itself in the liquid and not in
the gaseous state.
}\par}

\noindent
Based on this idea, Tisza [5,6] introduced a successful two-fluid 
model of the macroscopic behavior of Helium II; a similar model
was developed independently by Landau [7] on different grounds.

Here we shall report continued progress toward a quantitative 
microscopic understanding of the $\lambda$ transition in liquid $^4$He in 
terms of Bose-Einstein condensation of strongly interacting bosons.  
Carried out within the semi-analytic framework of correlated 
density-matrix theory, this effort aims at a concrete atomistic 
realization of London's hypothesis as further shaped by Feynman [8].  
In developing the theory, we have encountered a number of signposts 
erected by great physicists, suggesting that the path being 
followed is the right one.  As a preface to the presentation
of new results, it will be instructive to re-examine some of these
seminal insights.

Working in configuration space, the density matrix of a system of 
$N$ identical bosons can be written in the general form 
$$
W(\bfR,\bfR^\prime) = I^{-1} \Phi(\bfR) Q(\bfR,\bfR^\prime)
\Phi(\bfR^\prime) 
\, ,\eqno(1)
$$
where the coherence factor $\Phi(\bfR^\prime)$ becomes the ground-state 
wave function in the limit of zero temperature, $Q(\bfR,\bfR^\prime)$ 
is an incoherence factor that is non-separable in the configurations
$\bfR=(\bfr_1,\dots ,\bfr_N)$ and 
$\bfR^\prime=(\bfr^\prime_1,\dots ,\bfr^\prime_N)$ 
and describes the effects of real excitations, and $I^{-1}$ is a 
normalization constant.  It is convenient to split the incoherence 
factor into two parts, $Q(\bfR,\bfR^\prime) = Q_{\rm
coll}(\bfR,\bfR^\prime)
Q_{\rm qp}(\bfR,\bfR^\prime)$.  The factor $Q_{\rm coll}$ embodies the 
effects of real, thermal collective excitations (phonons and rotons),
while 
$Q_{\rm qp}$ accounts for the effects of real quasiparticle excitations
(or exchange correlations).  In the trial density matrix currently
employed, the factors $\Phi$, $Q_{\rm coll}$, and $Q_{\rm qp}$ are 
built from two-point functions.  In particular, $\Phi(\bfR)$ and 
$Q_{\rm coll}(\bfR,\bfR^\prime)$ are taken essentially as Jastrow 
products of appropriate dynamical and collective-thermal correlation 
functions [9-11] $\exp[u(|{\bf r}_i -{\bf r}_j|)/2]$ and 
$\exp \gamma(|{\bf r}_i -{\bf r}_j|)$, respectively, while 
$Q_{\rm qp}(\bfR,\bfR^\prime)$ is expressed in terms of a permanent 
of appropriate statistical correlation functions [12]
$\Gamma_{cc}(|{\bf r}_i -{\bf r}^\prime_j|)$.  Ultimately, the functions
$u$, $\gamma$, and $\Gamma_{cc}$ are to be determined by functional
minimization of the free energy.  A description of backflow effects
requires the inclusion of three-point functions in the coherence factor 
$\Phi(\bfR)$ and the incoherence factor $Q(\bfR,\bfR^\prime)$. 

Decomposition of the factors $\Phi(\bfR)$, $\Phi(\bfR^\prime)$, 
$Q_{\rm qp}(\bfR,\bfR^\prime)$, and $Q_{\rm coll}(\bfR,\bfR^\prime)$ into 
products of two-point functions (and, as needed, three- and higher-point 
functions) is motivated by the quantitative successes of the variational 
wave-function theory of the ground state of a strongly interacting 
Bose system [13].  In this theory, the ground-state wave function 
is written as a Jastrow product of pair correlation functions 
supplemented by a product of triplet correlation functions, and so 
on (as needed) to higher-body products in the Feenberg representation 
of the exact Bose ground state [11].

The treatment of $Q_{\rm qp}(\bfR,\bfR^\prime)$ in terms of a permanent
of two-point functions (the statistical correlation functions
$\Gamma_{cc}(|{\bf r}_i -{\bf r}^\prime_j|)$) exploits and extends the 
ideas of Feynman [8] and the formalism of Ziff, Uhlenbeck, and Kac [12].
For free bosons, which is the case studied by Ziff et al.,
$\Gamma_{cc}(r)$
has a Gaussian form.  Feynman used this form but inserted an effective
mass
in place of the bare mass.  

To quote Feynman's abstract,
{\par\narrower\noindent
{\sl
It is shown from first principles that, in spite of the large
interatomic forces, liquid $^4$He should exhibit a transition
analogous to the transition in an ideal Bose-Einstein gas.
The exact partition function is written as an integral over 
trajectories, using the space-time approach to quantum mechanics.
It is next argued that the motion of one atom through the others
is not opposed by a potential barrier because the others may move
out of the way.  This just increases the effective inertia of the
moving atom.  This permits a transition, but of the third order.
It is possible that a more complete analysis would show the 
transition implied by the simplified partition function is actually
like the experimental one.
}\par}

The strategy now being pursued within correlated density-matrix theory 
seeks to complete Feynman's program using a description that is more
flexible yet is rooted in the same physical picture in which
exchange effects are of paramount importance.  As the system
goes from the normal to the superfluid state, exchange loops
(or ``rings'' or ``cycles'') of macroscopic length come to dominate 
the behavior of the system; this feature is revealed in the
development of off-diagonal long-range order of the partial
distribution function (or correlation function) $G_{cc}(r)$
that characterizes exchange processes.  The path-integral 
Monte Carlo approach of Ceperley and Pollock [14,15] is also 
based on Feynman's ideas on liquid $^4$He, played out within Feynman's
space-time approach to quantum mechanics.  Quite naturally,
a similar picture emerges in this work -- involving a kind of 
percolation transition in which exchange paths can grow to 
macroscopic size.

The entropy corresponding to the chosen trial density matrix is 
determined using a replica ansatz (borrowed from the statistical 
physics of disordered systems), together with the so-called diagonal 
approximation, which implements a separability hypothesis analogous to 
that applied by Campbell and Feenberg [9] in the paired-phonon theory 
of the ground state.  The work of Ziff, Uhlenbeck, and Kac [12]
provides the pivotal clue for constructing a trial density matrix 
that is applicable in both normal and condensed phases and across 
the phase transition.

The presence of a {\it permanent} in the trial density matrix introduces
complications in the evaluation of spatial distribution functions and
structure functions and of the internal-energy and entropy components
of the free energy.  However, a powerful formalism for handling 
exchange correlations comes to the rescue:  the technical problems 
created by the permanent can be solved by appealing to (and extending) 
the methods of {\it Fermi} hypernetted-chain theory invented by 
Fantoni and Rosati [16] and Krotscheck and Ristig [17]. 
Fermi-hypernetted chain theory teaches us how to evaluate 
correlated expressions involving {\it determinants} of two-point 
statistical functions, and this knowledge is readily adapted to 
Bose statistics.  Complications also arise in the evaluation of 
the entropy using the replica ansatz, and these can be largely 
overcome with the aid of methods developed by Hiroike [18] 
for classical fluid mixtures.

The entropy, calculated in diagonal approximation, decomposes into 
independent contributions from phonon-roton excitations and from two 
kinds of quasiparticle excitations.  The internal energy also
decomposes into contributions from these three excitation branches, 
plus a correlation-energy term that becomes the ground-state energy 
at zero temperature.  Within the correlated density-matrix theory,
as explicated using the hypernetted-chain formalism, it is natural 
to interpret the two particle-type excitations as quasi{\it holes} and 
quasi{\it particles}.  The quasihole and quasiparticle dispersion 
relations are identical in the normal phase, where particle-hole 
symmetry is maintained, but they differ markedly in the condensed phase, 
where the exchange symmetry is broken.  The $\lambda$ transition 
therefore involves {\it two} order parameters: 
\item{(i)}
A condensation strength parameter $B_{cc}$ which vanishes in the
normal phase and measures the breaking of {\it gauge symmetry} in
the condensed phase (also occurring in the noninteracting Bose gas) 
\item{(ii)}
An exchange strength $M$ which also vanishes in the normal phase
and measures the violation of {\it exchange symmetry} below the $\lambda$ 
point (occurring only if the interactions are strong enough)  

\noindent
On physical grounds, these order parameters must be coupled, 
the simplest reasonable connection being a straight proportionality 
$M = a B_{cc}$, where the coupling parameter $a$ is temperature 
dependent [19]. 

The interpretation of the two quasiparticle branches as {\it quasihole}
and {\it quasiparticle} modes gains support from an idea voiced
by Felix Bloch to the effect that -- much as in a Fermi system -- there 
might exist quasiparticle and quasihole excitations in a system of
identical bosons.  Bloch's idea (unpublished, but aired at a 
Stanford seminar) resonates with Pines' phenomenological arguments 
[20] that {\it strongly} interacting Bose and Fermi liquids have 
much more in common than is normally supposed, a claim well documented
by experimental findings on the properties of their elementary
excitations.

The nature of the proposed exchange-symmetry-breaking phenomenon is
illuminated by the existence of a striking parallel with the behavior of 
a diamagnetic material in a magnetic field [19].  A particle-hole 
exchange field may be constructed as the sum of differences between 
the hole energies and the corresponding free single-particle energies, 
weighted by normalized differences between the occupation numbers of 
the two types of quasiparticle excitations.  Pursuing the electromagnetic 
analogy, a macroscopic polarization is induced by this field, which 
is negatively proportional to the field with a diamagnetic-susceptibility 
coefficient that depends on the occupations of the two quasiparticle 
branches.  This polarization effect acts to screen the exchange field 
in the condensed phase.  Indeed, in analogy with the Meissner effect, 
the condensed phase behaves like an ideal diamagnetic as $T$ approaches 
zero, completely expelling the exchange field from the system.  The 
screening leads to a drastic reduction in the particle-hole exchange 
contribution to the internal energy of the system.  This exchange
property was considered in detail in Ref.~[19], where the 
exchange-symmetry breaking was first proposed as an essential
aspect of the $\lambda$ transition in liquid $^4$He.

The diamagnetic analogy evokes a conviction expressed by John 
Bardeen [21] in recounting the development of superconductivity theory: 
{\par\narrower\noindent
{\sl
It seems to me that most of those who thought long and hard about
superconductivity prior to the discovery of the Meissner effect in
1933 never got over an inner feeling that the really fundamental
property of a superconductor is infinite conductivity or persistent
currents, and this colored the way they thought about the subject
in future years.  While an adequate theory must explain both aspects,
the diamagnetic approach has been the most fruitful in indicating
the nature of the superconducting state.\par}
}

\noindent
This deep insight sanctions an analogous description of the 
superfluid state of liquid $^4$He.

In advance of a complete functional optimization of the trial density 
matrix, a model calculation [19] has provided strong support for the new 
microscopic approach and especially for the importance of
exchange-symmetry 
breaking in the $\lambda$ transition.  A standard parameterized form 
was assumed for the dynamical correlation function (i.e.\ for the 
pseudopotential $u(r)$), while -- as in Feynman's treatment -- the 
statistical correlation function $\Gamma_{cc}(r)$ was determined through 
an effective-mass parameterization of the quasihole spectrum.  The 
collective (or phonon-roton) incoherence factor 
$Q_{\rm coll}(\bfR,\bfR^\prime)$,
which is not expected to exert a strong influence on the
transition, was set equal to unity.  A simple and plausible model was
introduced for the coupling parameter $a(T)$ connecting
the order parameter $M$ for exchange-symmetry violation to 
the order parameter $B_{cc}$ for gauge-symmetry breaking.  This
model requires, as empirical input, the ratio of the speeds of second
and first sound in the condensed phase.  

Within the semi-microscopic/semi-empirical model so delineated, exchange-
symmetry breaking was found to reduce the condensation temperature 
of liquid $^4$He from the value 3.2 K corresponding to Bose-Einstein
condensation of a noninteracting Bose gas to a value 2.2 K very close 
to the experimentally measured $\lambda$ transition temperature.
This prediction is based entirely on the extrapolated behavior of 
the hole chemical potential in the normal phase; it is consistent 
with the behavior of the model in the condensed phase but does not
depend on the empirical input for the sound-speed ratio.  The 
prediction for the specific heat shows an apparently divergent behavior 
at 2.2 K and qualitatively -- though not quantitatively -- reproduces 
the characteristic $\lambda$ shape of the experimental curve.  The 
latter result contrasts with that of Feynman's treatment, where 
an approximate and incomplete analysis of the geometry of exchange 
processes led to a {\it continuous} specific heat, as in the 
noninteracting Bose gas.

In this paper we continue to build upon the correlated density-matrix 
theory of strongly correlated Bose systems, essentially in the form 
designed and developed in Refs.~[22-27].  The trial density matrix 
for liquid $^4$He and the consequent structure of the energy and 
entropy (the latter evaluated in diagonal approximation) are made 
more explicit in Sec.\ 2.  The Euler-Lagrange equations for the 
optimal density matrix without backflow, expressed in convenient form 
(as presented in Sec.\ 3), are solved numerically.  The results 
of the full optimization are examined in detail in Sec. 4.  (Additional 
information is furnished by Ref.~[28].)  Remaining obstacles to a 
full microscopic understanding of the $\lambda$ transition are discussed
in
the final section.  Detailed formulas of the theory are relegated
to an appendix.

\vskip 28 truept
\centerline{\bf {2.  CORRELATED TRIAL DENSITY MATRIX AND FREE ENERGY}}
\vskip 12 truept

The theory begins with suitable choices for the coherence and 
incoherence factors appearing in Eq.~(1):
\item{(i)} A wave function of Jastrow form
$$
\Phi(\bfR)=\exp\left\{{1\over2 }\sum^N_{i<j}u(|\bfr_i-\bfr_j|)\right\} \, 
,\eqno(2)
$$
\item{(ii)} An incoherence factor $Q=Q_{\rm coll}Q_{\rm qp}$ that 
takes account of real collective excitations (phonons, rotons)
through
$$
Q_{\rm coll}(\bfR,\bfR')=\exp\sum_{i,j}^N\left\{\gamma\left(|\bfr_i
-\bfr'_j|\right)-{1\over 2}\gamma\left(|\bfr_i-\bfr_j|\right)-{1\over
2}\gamma\left(|\bfr'_i-\bfr'_j|\right)\right\}\ , \eqno(3) 
$$ 
and incorporates the effects of quasiparticle excitations via
$$
Q_{\rm qp}(\bfR,\bfR')={1\over A^N}{1\over
2\pi{\rm i}}\oint\!{dz\over z}\,{\rm
e}^{A/z}\perm\limits_{i,j}\bigl[\gamcc\left(|\bfr_i-\bfr'_j|\right)
+\bcc z\bigr]\ .\eqno(4) 
$$ 
The contour in Eq.~(4) encircles the origin in the counterclockwise sense, 
and the quantity $A$ is an integral involving $\Gamma_{cc}$
and the $dd$ and $dc$ components of the radial distribution function
[19]. The condensation strength $\bcc$ serves as the primary order
parameter that monitors the breaking of gauge symmetry as one enters
the Bose-Einstein condensed phase, being restricted by $0<\bcc\le 1$.
In the normal phase where $\bcc\equiv0$, the integration in expression (4)
may be carried out analytically and the incoherence factor 
$Q_{\rm qp}(\bfR,\bfR^\prime)$ reduces simply to the permanent
of the statistical functions $\Gamma_{cc}(|\bfr_i-\bfr'_j|)$ alone.

The free-energy functional associated with the trial density matrix
specified by Eqs.~(2)--(4) has been constructed explicitly [19],
with the following general results. The internal energy of the $N$-boson 
system at temperature $T$ is the sum of a correlation energy $E_c$ 
(which coincides with the expected ground-state 
energy at zero temperature) and separate contributions from collective, 
quasihole, and quasiparticle excitation branches,
$$
E=E_c+ E_{\rm coll} + E_{\rm qh}+ E_{\rm qp}\, .\eqno(5) 
$$ 
The energy addends $E_{\rm coll}$, $E_{\rm qh}$, $E_{\rm qp}$ account,
respectively, for the individual renormalized energy contributions 
$\epsilon_{\rm coll}(k)$, $\epsilon_{cc}(k)$, and 
$\epsilon_{0}(k)= \hbar^2 k^{2}/2m$ from phonons (or rotons), quasiholes,
and 
quasiparticles of all momenta $\hbar\bfk$,
$$\eqalignno{
E_{\rm coll} &=  \sum_\bfk\epsilon_{\rm coll}(k)n(k)\, ,&(6)\cr
E_{\rm qh} &={1\over2}\sum_\bfk\epsilon_{cc}(k)n_{cc}(k)\, ,&(7)\cr
E_{\rm qp} &={1\over2}\sum_\bfk\epsilon_{o}(k)n_{c}(k)\, .&(8)\cr }
$$
In turn, the functions $n(k),\ n_{cc}(k)$, and $n_c(k)$ are the 
respective thermally averaged occupation numbers of these three kinds
of excitations, at wave vector $\bfk$. 
The corresponding excitation energies $\omega(k)$, 
$\omega_{cc}(k)$, and $\omega_c(k)$ are defined through the Bose 
distribution function
$n(k)=\left[\exp\beta\omega (k)-1\right]^{-1}$ and through the analogous
Bose distributions $n_{cc}(k)$ and $n_c(k)$.  

In the {\it diagonal} approximation [24,25], the three types of 
excitations also contribute additively to the total entropy $S_e$,
$$
S_e=S_e^{({\rm coll})}+S_e^{(cc)}+S_e^{(c)}\, ,\eqno(9)
$$
where 
$$
\eqalignno{
  S_e^{({\rm coll})} &= k_B\sum_\bfk\Bigl\{\bigl[
1+n(k)\bigr]\ln\bigl[ 1+n(k)\bigr]-n(k)\ln
n(k)\Bigr\}\, ,&(10)\cr
S_e^{(cc)} &= {1\over 2}k_B\sum_\bfk\Bigl\{\bigl[
1+n_{cc}(k)\bigr]\ln\bigl[ 1+n_{cc}(k)\bigr]-n_{cc}(k)\ln
n_{cc}(k)\Bigr\}\, ,&(11)\cr}
$$
and $S_e^{(c)}$ is obtained from Eq.~(11) by replacing $n_{cc}(k)$ 
with $n_c(k)$.

The energies $E_c$, $\epsilon_{\rm coll}(k)$, and $\epsilon_{cc}(k)$ and
the distributions $n(k)$, $n_{cc}(k)$, and $n_c(k)$ are to be
regarded as functionals of the statistical function $\gamcc (r)$ 
and the pseudopotentials $u(r)$ and $\gamma(r)$.   In the condensed 
phase, they are also functions of the condensation strength $\bcc$.  
Explicit expressions for these quantities are provided in the Appendix.

The two excitation branches having Bose distribution functions
$n_{cc}(k)$ and $n_c(k)$ and respective energies $\omega_{cc}(k)$ 
and $\omega_c(k)$ are characterized at small wave number $k$ by 
a quadratic dispersion law or/and an energy gap.  In the normal 
phase where $\bcc\equiv 0$ these two branches merge, i.e., 
$n_{cc}(k)=n_c(k)$ and $\omega_{cc}(k)=\omega_c(k)$ 
for any $k\ge 0$. However, in the condensed phase where $\bcc>0$,
this degeneracy may be lifted in the presence of interactions between
the boson constituents. The two branches then follow different dispersion 
laws.  In the spirit of Felix Bloch's conjecture as enunciated in the
introduction, we view the $cc$ and $c$ branches as involving quasihole 
and quasiparticle excitations, respectively [19].  According to this
Bloch interpretation, the symmetry between particle and hole excitations 
is broken in the condensed phase of a strongly interacting Bose system.  
Further, as sketched in the introduction, the exchange-symmetry-breaking 
phenomenon can be given a formal expression by developing an analogy 
with the behavior of a diamagnetic material in a magnetic field.  The
counterpart of the Meissner effect (expulsion of the exchange field)
has been studied numerically in Ref.~[19] within a model in which simple 
parameterized forms are assumed for the dynamical correlations and
for the hole spectrum that determines the statistical correlations. 

To proceed beyond the model of Ref.~[19] toward a systematic microscopic
treatment, we now turn to the problem of full functional optimization of 
the Helmholtz free energy.  Due to the formal complexity of this problem, 
the present implementation is limited to the choice (2)--(4) for the
trial density matrix, implying the omission of three- and higher-body
correlation functions and hence neglect of what are traditionally
called ``backflow'' effects.

\vskip 50 truept
\centerline{{\bf 3.  FUNCTIONAL OPTIMIZATION OF THE FREE ENERGY}}
\vskip 12 truept

The Delbr\"uck--Gibbs--Moliere minimum principle for the Helmholtz 
free energy [23] is employed to determine the optimal density matrix 
(1) within the class of trial functions defined by Eqs.~(2)--(4).  
Variation of the functional $F=E-TS_e$ constructed from the results 
(5)--(11) is, however, constrained by a particle-number (or
particle-conservation) sum rule.  This sum rule takes the form [25]
$$
N= \bcc N_{o}+\sum_\bfk n_{cc}(k) \eqno(12)
$$
with
$$
N_{o}=N +\sum_\bfk\gamcc
(k)\left[S_{dc}(k)+S_{cc}^{(1)}(k)\right]\eqno(13)
$$
in the condensed phase where $0<\bcc\le 1$; it of course lacks the
$N_o B_{cc}$ term in the normal phase. 
The functions $S_{dc}(k)$ and $S_{cc}^{(1)}(k)$ are components of 
the structure function $S(k)$ and the exchange function 
$S_{cc}(k)$, respectively [19].  Coupling the condition (12) 
to the free energy expression with a Lagrange parameter $\lambda$, 
the modified functional $F_\lambda$ is varied with respect to the 
input functions $u(r)$, $\gamcc (r)$, and $\gamma (r)$ (or 
their Fourier transforms $u(k)$, $\gamcc (k)$, and $\gamma (k)$). The 
Euler-Lagrange equations
$$
{\delta F_\lambda\over\delta u(k)}=0\, ,\quad
{\delta F_\lambda\over\delta \gamcc (k)}=0\, , \quad
{\delta F_\lambda\over\delta \gamma (k)}=0 \eqno(14)
$$
then determine the optimal pseudopotential, quasihole/quasiparticle
spectra,
and phonon/roton spectrum.

Explicit expressions for the variational derivatives in Eqs.~(14) 
(again, in the absence of backflow) may be found in Ref.~[28].  The 
results can be recast in the form of a generalized paired-phonon equation 
[9,10] 
$$
\tbul{S}(k)+{1\over 2}\epsilon_0(k)\bigl[S(k)-1\bigr]=  D(k)\eqno(15)
$$
for the radial distribution function or the associated static 
structure function $S(k)$, an analogous equation 
$$
\tbul{S}_{cc}(k)+{1\over 2}\epsilon_0(k) S_{cc}(k)= D_{cc}(k) \eqno(16)
$$
for the statistical exchange function $G_{cc}(r)$ or the corresponding 
Fourier transform $S_{cc}(k)$, and a Feynman equation [29,22] 
$$
\omega(k)={\epsilon_0(k)\over S(k)}\coth\left[{\beta\over
2}\omega(k)\right]\eqno(17)
$$
for the collective excitations with the dispersion law 
$\omega=\omega(k)$ at temperature $T=(k_{B}\beta)^{-1}$.
Equations~(15) and (16) involve the generalized structure functions
$\tbul{S}(k)$ 
and $\tbul{S}_{cc}(k)$ along with the functions $D(k)$ and $D_{cc}(k)$. 
The Appendix and Ref.~[28] furnish detailed formulas and relations for 
these quantities.

To solve the paired-phonon equation (15), one customarily expresses it
-- equivalently -- as a generalized Bogoliubov equation
$$
\omega^2(k)=\epsilon_0(k)\bigl[\epsilon_0(k)+2v_{p-h}(k)-2v_{\rm
coll}^\ast(k)
\bigr]
\, , \eqno(18)
$$
where the collective potential is given by $v_{\rm
coll}^\ast(k)=-2\epsilon_0(k)n(k)(n(k)+1)/S^2(k)$ and
the particle-hole potential $v_{p-h}(k)$ is taken from Ref.~[28].

It is instructive to note that the correlated density-matrix theory may 
be fruitfully viewed as a pair-density energy-functional approach~[30].
The theory provides for systematic construction of a universal energy 
functional $E = E[G(r), G_{cc}(r),\bcc, \rho]$ which depends on one-body 
quantities (particle density $\rho$ and condensation strength $\bcc$) 
{\it and} on two-body densities (the radial distribution function $G(r)$ 
and the statistical exchange function $G_{cc}(r)$).  Similarly, the 
internal energy, the entropy, and the Helmholtz free energy are all 
functionals of these one- and two-body quantities.  Correlated 
density-matrix theory may be therefore interpreted as a generalization 
of the Kohn-Sham approach~[31] that explicitly acknowledges the role 
played by the dynamical and statistical correlations present in 
the many-body system.

We further observe that the generalized paired-phonon equation (15) may be 
transformed into a non-linear Schr\"odinger equation~[28,30]. This
equation has the form~[28]
$$
\left\{-{\hbar^{2}\over m} \Delta + v(r) + w(r)+ w_{\rm coll}(r) + w_{\rm
qp}
(r)\right\}\sqrt{g(r)}=0 \, , \eqno(19)
$$
where $v(r)$ is the bare $^4$He--$^4$He interaction and the induced
potentials 
$w(r)$, $w_{\rm coll}(r)$, and $w_{\rm qp}(r)$ are functionals 
of $G(r)$ and $G_{cc}(r)$ and functions of particle density $\rho$ and 
temperature $T$.
\vskip 28 truept
\centerline{{\bf 4.  NUMERICAL RESULTS AND DISCUSSION}}
\vskip 12 truept

Evaluation of the various components of the structure functions $S(k)$ 
and $\tbul{S}(k)$ and the statistical functions $S_{cc}(k)$ and 
$\tbul{S}_{cc}(k)$ is carried out with the hypernetted-chain 
techniques [25,32], neglecting elementary diagrams.  The Bogoliubov 
equation (18) is employed to solve the Euler-Lagrange equations 
efficiently by means of a suitable iteration procedure.  In the normal 
phase, the sum rule (12) serves to determine the chemical potential 
$\mu_{cc} =-\omega_{cc}(0)$ of the quasiholes (and quasiparticles). 
The condensation strength characterizing the condensed phase 
is obtained from relation (12) in conjunction with the assumed 
optimization condition 
$
\partial F_\lambda/\partial \bcc =0.
$
It must be noted, however, that the latter prescription is not 
entirely adequate; in future work it should be replaced by a 
more appropriate renormalized Hartree equation. The HFDHE2 Aziz 
potential [33] is adopted for the interaction between the $^4$He 
atoms.  The $^4$He atomic bosons are confined in a cubic box at 
a uniform density given by the experimental saturation density 
$\rho(T)$ of the liquid at temperature $T$, starting with the value 
$\rho=0.02185\ {\rm \AA}^{-3}$ at zero temperature.  Reference~[28]
contains further information on the numerical procedures applied.
Optimal versions of the various quantities have been calculated
in the temperature range $T=[0\ {\rm K},5.5\ {\rm K}]$.  
Figures~1--7 display a selection of the numerical results.
\midinsert
\input psfig.sty
\centerline{
{\psfig{figure=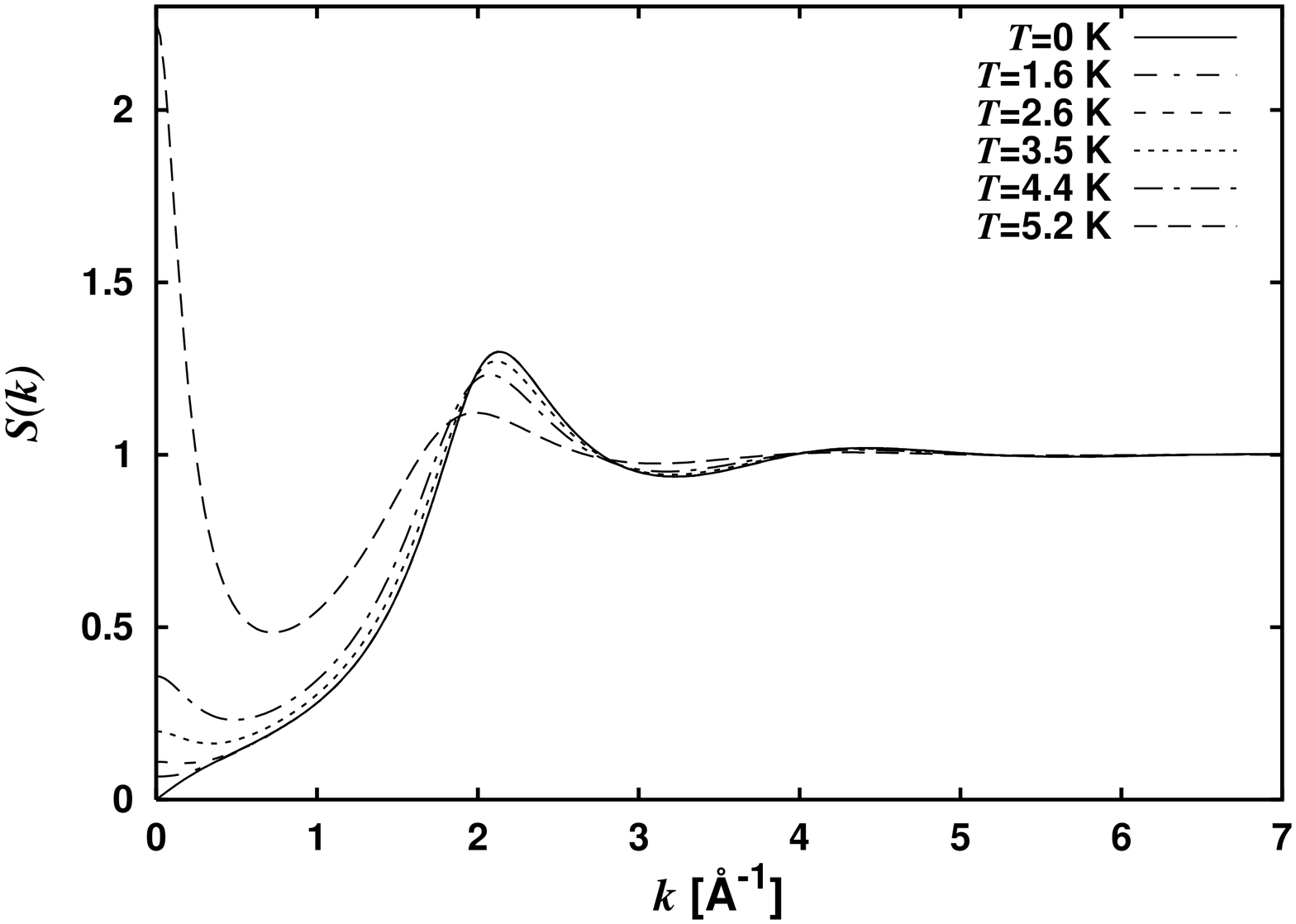,angle=270,width=14truecm,height=9truecm}}
}
\vskip 0.4truecm 
\noindent
{\bf Figure 1.} 
The optimal structure function $S(k)$ at various temperatures and at
the experimental saturation density.
\endinsert
\smallskip
Figure 1 presents the results for the structure function at several
temperatures.  As expected, the calculated data look very similar to 
earlier results [23], demonstrating that the exchange correlations 
(while certainly important for other quantities) affect 
the spatial structure of the $^4$He liquid only in its details. 
The variation of the low-$k$ behavior of the structure function reflects
the decrease of the velocity of first sound with increasing temperature,
as has been discussed in detail in Ref.~[23]. The maximum of $S(k)$ 
at about $k\simeq 2\ {\rm \AA}^{-1}$ shows a monotonic decrease 
with increasing temperature.  This predicted behavior is in 
accord with experimental measurements in the normal phase, but it
conflicts with experiment in the condensed phase, where the maximum 
is found to increase with temperature.  The generalized Feynman 
relation (17) yields theoretical results for the excitation 
energies of the collective phonon/roton branch.  The optimal data 
plotted in Fig.~2 shows the familiar dependence of $\omega(k)$ on 
wave number $k$: the linear dispersion law for phonons at low
momenta and the typical roton minimum at $k\simeq 2\ {\rm \AA}$. 
However, since backflow contributions have been sacrificed in favor of 
a closer examination of the effects of exchange correlations,
the roton minimum is more than twice what is seen experimentally.
Consequently, the number of elementary excitations increases only slowly 
with increasing temperature.  To examine more closely the implications
of this behavior, let us assume for the moment that Feynman's 
treatment [34] is applicable and that the superfluid density 
$\rho_s$ of the $^4$He liquid is approximately described by the expression
$$
\rho_s=1-{2\beta\over 3N}\sum_\bfk\epsilon_0(k)\bigl[1+n(k)\bigr]n(k)\, .
\eqno(20)
$$
If the distribution $n(k)$ takes small values, then the excitation energy
$\omega(k)$ is large and the sum in Eq.~(20) is small;  
therefore the superfluid density is close to unity.  Accordingly,
high temperatures are needed to make $\rho_s$ 
become small.  This rough line of reasoning already indicates that the 
current realization of density-matrix theory will predict
a value for the normal-to-superfluid transition temperature 
(i.e., the temperature $T_\lambda$ of the $\lambda$ transition) 
that is significantly higher than the experimental value.
\pageinsert
\input psfig.sty
\centerline{
{\psfig{figure=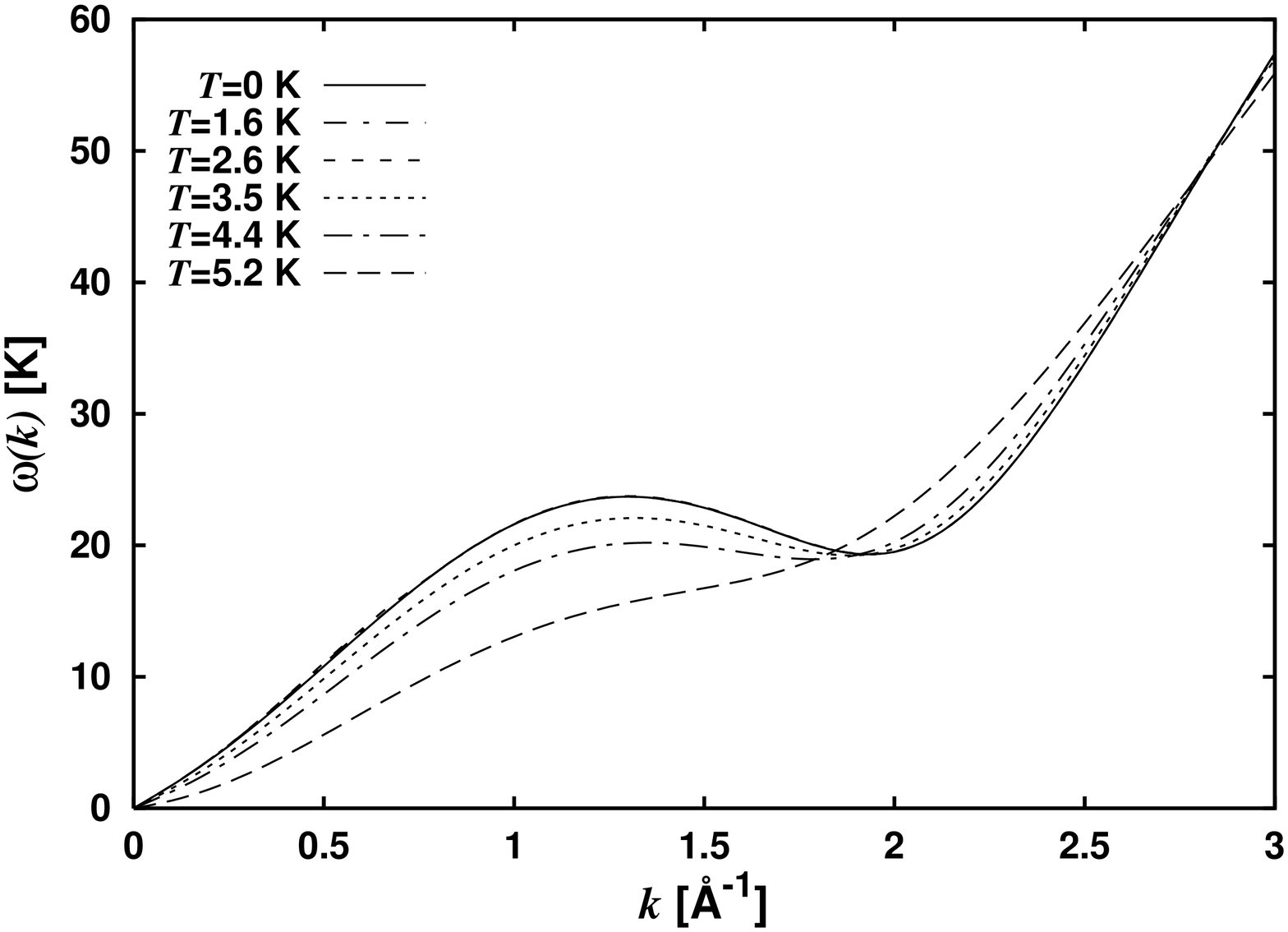,angle=270,width=14truecm,height=9truecm}}
}
\vskip 0.4truecm 
\noindent
{\bf Figure 2.}
The optimal phonon/roton excitation energy $\omega (k)$ as a function of
wave number $k$ at various temperatures and at the experimental saturation 
density.
\vskip 18truept
\vskip 1truecm 
\input psfig.sty
\centerline{
{\psfig{figure=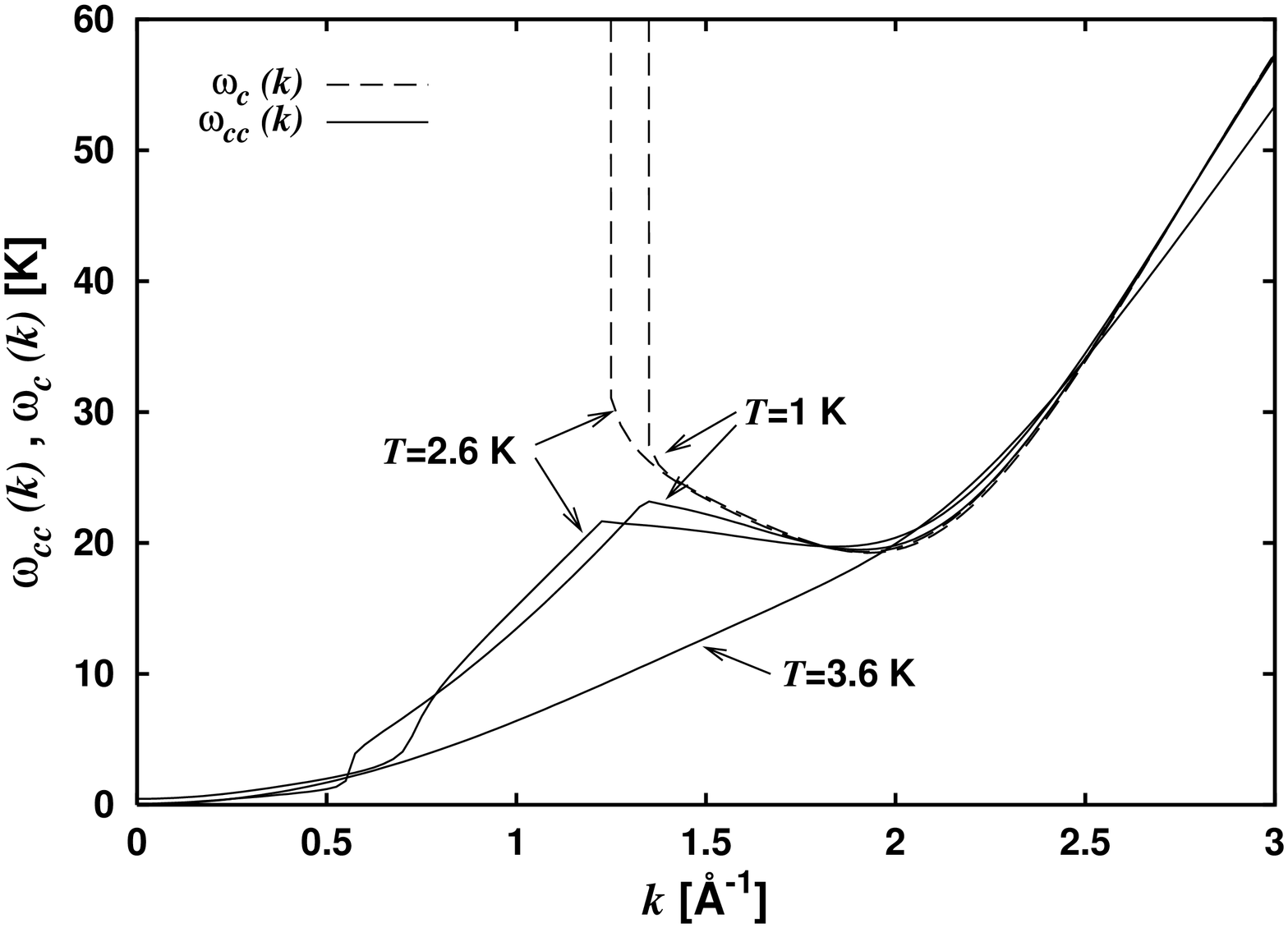,angle=270,width=14truecm,height=9truecm}}
}
\vskip 0.4truecm 
\noindent
{\bf Figure 3.}
The optimal quasihole energy $\omega_{cc}(k)$ and quasiparticle energy
$\omega_c (k)$ as functions of wave number $k$ at various temperatures and
at the experimental saturation density.
\endinsert

This feature of the calculation is clearly evident in the results 
for the respective excitation spectra $\omega_{cc}(k)$ and $\omega_c(k)$ 
of the quasiholes and the quasiparticles.  Figure~3 shows their 
dispersion curves for particle density $\rho=0.02185\ {\rm \AA}^{-3}$ 
at three different temperatures. The excitation energies are seen
to exhibit a minimum around the roton minimum for 
temperatures below $T< 3.1$ K.  In contrast to the case of quasiholes, 
the quasiparticle spectrum $\omega_c (k)$ is restricted to wave numbers 
$k>k_0$, excitations being forbidden at smaller wave numbers.  
The threshold $k_0$ for the onset of quasiparticle excitations 
decreases slowly with $T$ at very low $T$, but it decreases very
rapidly for $T\ge 2.8$~K and vanishes at $T\simeq 3.4$~K.  At higher 
temperatures, the quasiparticle and quasihole dispersion curves 
coincide and show a monotonic increase with increasing wave number $k$.

The low-temperature behavior of the excitation energies can
be extracted analytically from the Euler-Lagrange equations. 
One readily finds
$$
\omega_{cc}(k)=2\bigl[\epsilon_0(k)-\lambda\bigr]\ ,
\quad k<k_0\, ,\eqno(21)
$$
$$
\omega_c(k)=\omega_{cc}(k)=\bigl[\epsilon_0(k)-\lambda\bigr]S^{-1}(k)\ ,
\quad k>k_0\, ,\eqno(22)
$$
where the threshold $k_{0}$ is determined by $2 S(k_{0}) = 1$.
The chemical potential of the quasiholes is therefore
$\mu_{cc}=-\omega_{cc}(0)=2\lambda$ in the limit $T\rightarrow 0$.  

The numerical results for the optimal potential $\mu_{cc}$ in the
temperature range $0\le T\le 5.5$~K are plotted in Fig.~4.  As the
temperature is lowered from the ``large''-$T$ side, $\mu_{cc}(T)\le 0$ 
increases until it vanishes at a predicted transition temperature 
$T_\lambda\simeq 3.4$~K.  In the temperature range $T\ge T_\lambda$, 
the liquid is in the normal phase where $\bcc =0$.  For $\bcc>0$, 
there is effectively a potential barrier for quasiholes, formed 
by the sudden decrease and subsequent slow rise of $\mu_{cc}$ as 
the temperature is decreased from $T\le 3$~K.
As $T$ goes to zero, the hole chemical potential 
vanishes again and we have $\mu_{cc}(0)=2\lambda=0$.  However,
this low-$T$ behavior of $\mu_{cc}$ is an artifact of 
the variational condition $\partial F_\lambda/\partial \bcc =0$ 
that we have imposed on the $B_{cc}$ parameter.  We expect that 
an improved Hartree condition on the condensation strength will
produce a quasihole chemical potential that increases monotonically 
with temperature in the condensed phase and 
vanishes at the $\lambda$ transition point.
\pageinsert
\input psfig.sty
\centerline{
{\psfig{figure=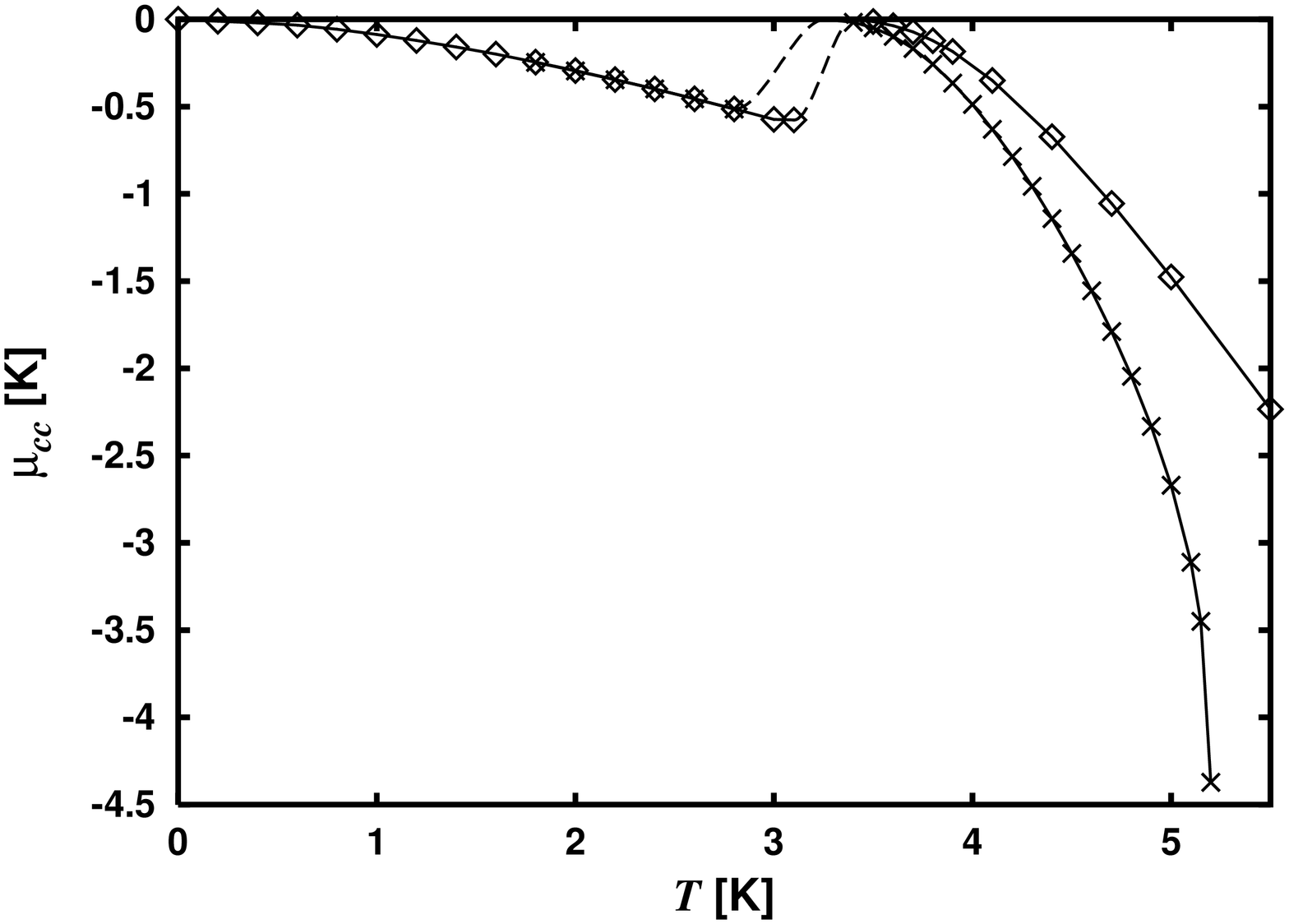,angle=270,width=14truecm,height=9truecm}}
}
\vskip 0.4truecm 
\noindent
{\bf Figure 4.}
The optimized quasihole chemical potential $\mu_{cc}$ versus temperature
at
density $\rho= 0.02185\ {\rm \AA}^{-3}$ (diamonds) and at the experimental 
saturation density (crosses).
\vskip 18truept
\input psfig.sty
\centerline{
{\psfig{figure=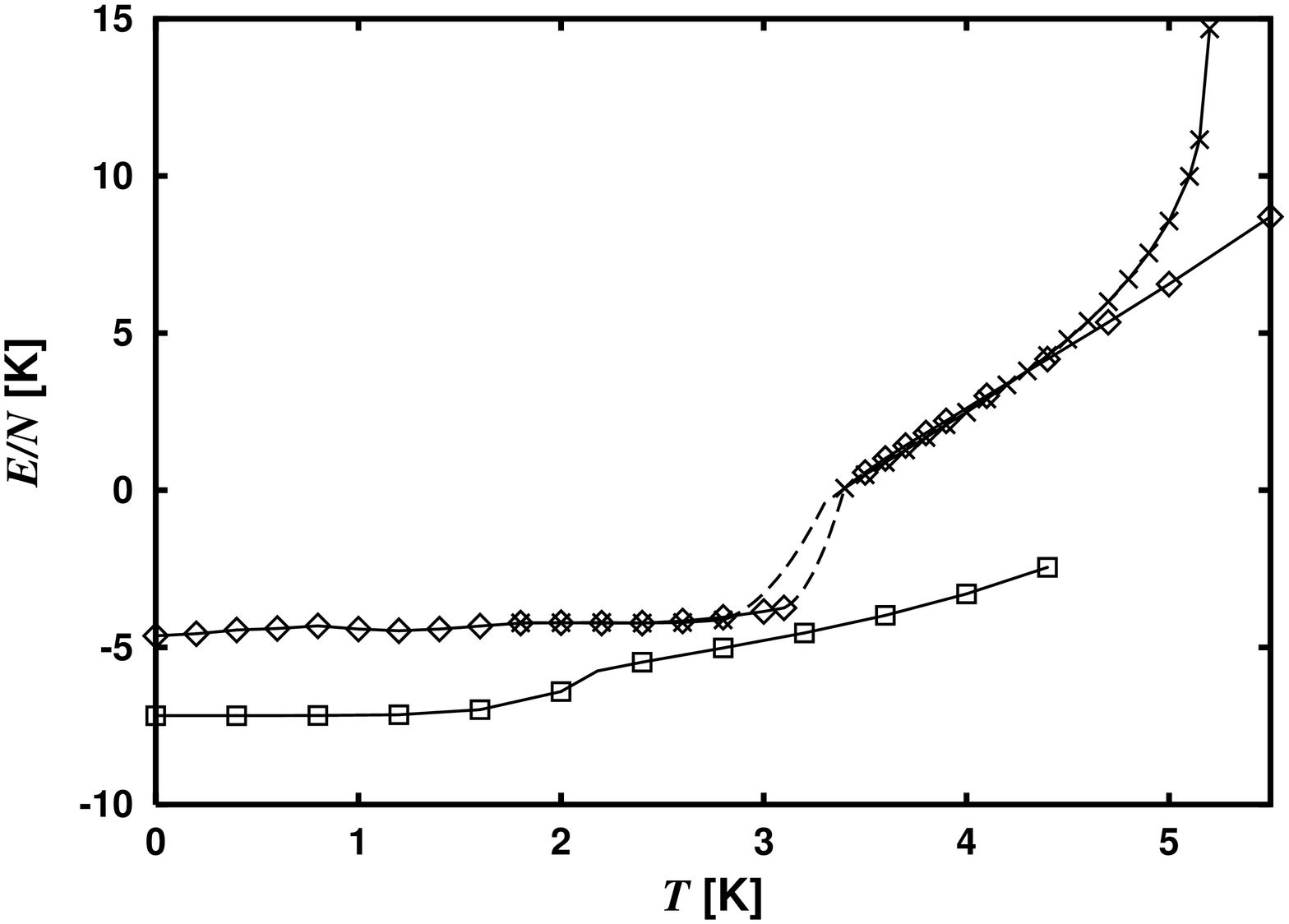,angle=270,width=14truecm,height=9truecm}}
}
\vskip 0.4truecm 
\noindent
{\bf Figure 5.}
Optimal theoretical results for the energy per particle as a function
of temperature $T$ at $\rho=0.02185\ {\rm \AA}^{-3}$ (diamonds) and at
saturation density (crosses). The squares mark experimental data taken 
from Ref.~[35].
\endinsert
\smallskip

The numerical predictions for the excitation energies $\omega_{cc}(k)$ 
and $\omega_c(k)$ themselves are very large, implying 
that the associated Bose distributions $n_{cc}(k)$ and $n_c(k)$ are 
unrealistically small.  The particle-number sum rule 
(12) then leads inevitably to a large critical temperature 
$T_\lambda\simeq 3.4$~K, even larger than for a Bose gas of 
non-interacting $^4$He atoms and much too high compared with 
the experimental $\lambda$-point, $T_\lambda=2.18$~K. 
It would seem that the present ansatz for the density matrix, devoid
of backflow, is not flexible enough to produce a sufficiently large
effective mass of the quasiholes in the medium and a concomitant 
reduction of the excitation energy $\omega_{cc}(k)$.  This perspective
is in harmony with Feynman's argument that the motion of one helium 
atom through the others in the liquid increases the effective inertia 
of the moving atom [8].  The parameterized model of Ref.~[19] 
also lends support to this assessment: it involves
a large quasihole effective mass $m^\ast\simeq 1.5m$ and 
yields a theoretical transition temperature $T_\lambda\simeq 2.2$~K, 
in quantitative agreement with results derived from the present
formulation by simulating the effects of backflow (see below).

Figure~5 plots the calculated optimal data for the internal energy per 
particle $E/N$ as a function of temperature at the fixed density 
$\rho=0.02185\ {\rm \AA}^{-3}$ and at the ($T$-dependent) experimental 
saturation density. In the temperature range $0\le T\le 2$~K, the 
theoretical predictions (at saturation density) lie higher than the 
experimental results by about 2.5~K.  Numerical calculations within 
correlated-basis-function theory [36,37] as well as 
variational Monte-Carlo studies [38] at zero temperature strongly 
suggest that the energy excess is due in large part to the omission 
of triplet pseudopotentials $u(\bfr_i,\bfr_j,\bfr_k)$ from
ansatz (2).  At zero temperature, these terms essentially 
account for the energetic effect of backflow.  As expected, the high 
energy per particle obtained in the present treatment is accompanied 
by a large theoretical value for the transition temperature, 
$T_\lambda\simeq 3.4$~K.

The Bose-Einstein-condensed phase and the associated violation 
of gauge symmetry may be characterized by the primary order parameter
$\bcc$, which we have termed the condensation strength, or alternatively 
by the non-zero condensate fraction $n_0(T)$.  In the temperature 
interval $0\le T<3$~K, the $\bcc$ order parameter depends rather 
weakly on $T$, decreasing linearly from unity at $T=0$ to about 
$0.8$ at $T=3$~K. For slightly higher temperatures, a drastic 
reduction sets in and $\bcc$ vanishes 
at $T_\lambda\simeq 3.4$~K.  In this narrow region of rapid falloff
of $\bcc$, the behavior of the solutions of the Euler-Lagrange equations 
depends sensitively on the properties of the quasihole and quasiparticle 
excitation branches, which tends to destabilize the iteration process
we have employed (for details, see Ref.~[28]).
\topinsert
\input psfig.sty
\centerline{
{\psfig{figure=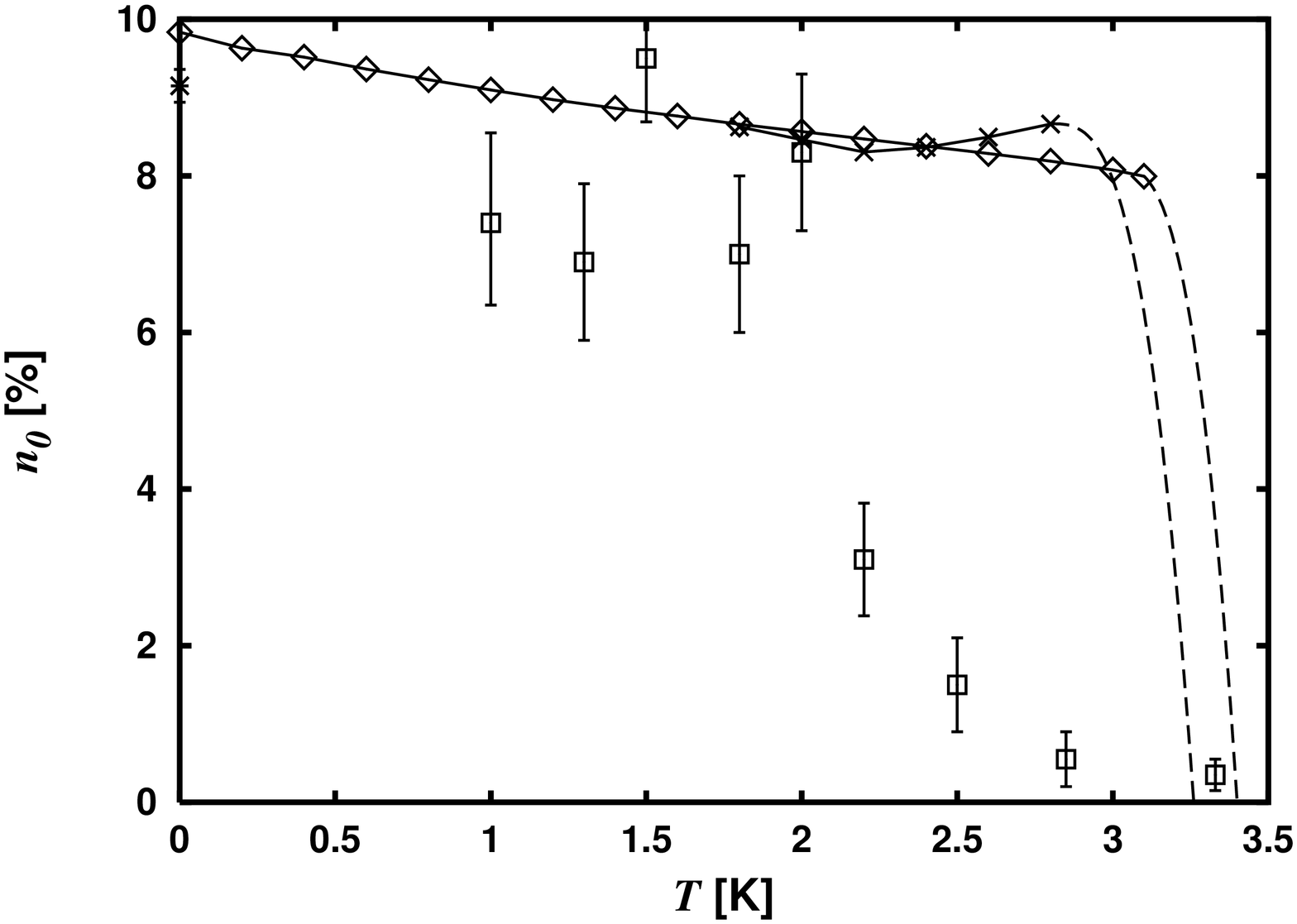,angle=270,width=14truecm,height=9truecm}}
}
\vskip 0.4truecm 
\noindent
{\bf Figure 6.} 
Optimal theoretical results for the condensate fraction $n_0$ versus 
temperature at density $\rho=0.02185\ {\rm \AA}^{-3}$ (diamonds) and at 
experimental saturation density (crosses).  Shown for comparison are data
from path-integral Monte Carlo [14] and Green's function Monte Carlo
[41] studies (denoted respectively by squares with error bars and
by the asterisk).
\endinsert
\smallskip

The condensate fraction $n_0(T)$ has been calculated within the 
hypernetted-chain formalism of Ref.~[39]. The numerical results are 
plotted in Fig.~6, where they are compared with the path-integral 
Monte Carlo results of Ceperley and Pollock [14].  At zero temperature, 
we reproduce the value $n_0(0)\simeq 9.8$\% obtained in an earlier 
application of correlated density-matrix theory and reported in Ref.~[40].  
This value is satisfactorily close to the estimate given by the
Green's function Monte Carlo approach [41], $n_0(0)\simeq 9.0\pm 0.3\%$. 
With increasing temperature below 2 K, the optimal results for $n_0(T)$
show a gradual linear decrease similar to that suggested by the 
path-integral Monte-Carlo data.  However, the expected precipitous
decline toward a vanishing condensate fraction is not seen until
a rather large temperature is reached, $T\simeq 3.2$~K.  This behavior, 
again pointing to a normal-superfluid transition temperature that is
too high, is consistent with the above findings for the excitation 
spectra, internal energy, and condensation strength.  The results
of the current treatment for other observable quantities, including 
the entropy and velocity of sound, are discussed in Ref.~[28]. 

To incorporate backflow effects into the present framework, we
must generalize the formulae (2)--(4) defining the choice of trial 
density matrix.  This may be done by supplementing the 
pseudopotentials $u(r)$ and $\gamma(r)$ by 
triplet (or three-point) factors and by generalizing the permanent in 
expression (4). Alternatively, we may introduce higher-order correlations 
through the device of shadow wave functions [42,43]. 
Either option is both formally and computationally demanding,
especially since elementary-diagram contributions should be included
for consistency.  In advance of such a program, it is useful to
adopt a more phenomenological approach: we shall assume that 
``backflow'' has the principal effect of renormalizing the boson
mass $m$ appearing in the collective and quasiparticle and 
quasihole terms of the energy (5).  Restricting ourselves to a
treatment of the normal phase, we accordingly modify the functionals 
(6)--(8), derived from first principles, to read
$$
\eqalignno{
E_{\rm coll} &=  \sum_\bfk{m\over m^\ast_{\rm coll}}\epsilon_{\rm coll}
(k)n(k)\, ,&(23)\cr
E_{\rm qh} = E_{\rm qp}  &={1\over4}\sum_\bfk {m\over m^\ast_{\rm qp}}
\bigl[\epsilon_{cc}(k)+ \epsilon_{0}(k)\bigr]n_{cc}(k)\, . &(24) \cr}
$$  
In general, the effective masses $m^\ast_{\rm coll}$ and $m^\ast_{\rm qp}$
depend on temperature, density, and wave number and differ
from each other.

As anticipated, a revised optimization based on the modified 
functionals (23) and (24) does indeed reveal a strong sensitivity 
of the theoretical transition temperature $T_{\lambda}$ to variations 
in the effective mass.  Figure~7 collects numerical results 
for the optimal chemical potential $\mu_{cc}$ of quasiparticles/quasiholes 
in the normal phase at density $\rho=0.02185 {\rm \AA}^{-3}$,
for a selection of renormalized masses $m^\ast_{\rm qp}$.  
The theoretical value for $T_{\lambda}$ is very close to the 
experimental result of $2.18$ K, if the choice $m^\ast_{\rm qp}\approx 
1.6m$ is adopted.  
The results obtained for the connection between the predicted 
$T_{\lambda}$ and the effective quasiparticle mass are essentially
independent of the value chosen for the mass parameter $m^\ast_{\rm
coll}$.
We note further that a judiciously tailored temperature-dependent 
$m^\ast_{\rm qp}(T)$ could be used to match the theoretical results 
for the specific heat to the experimental data in the normal phase.
\topinsert
\input psfig.sty
\centerline{
{\psfig{figure=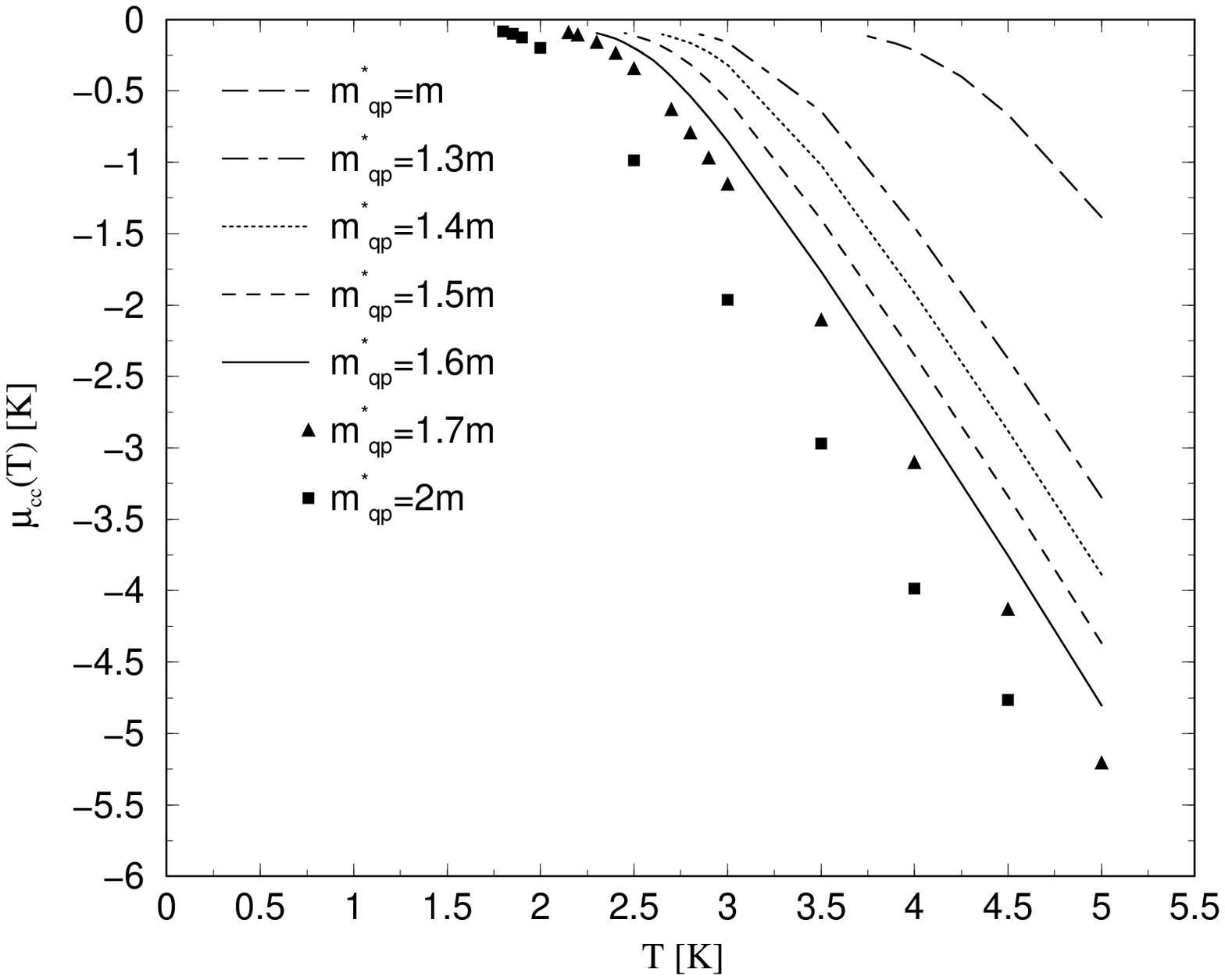,angle=0,width=14truecm,height=9truecm}}
}
\vskip 0.4truecm 
\noindent
{\bf Figure 7.}
The chemical potential $\mu_{cc}(T)$ of the quasiparticles (quasiholes) 
in the normal phase as function of temperature $T$  at density 
$\rho=0.02185 {\rm \AA}^{-3}$ for different values of  the effective 
mass $m^\ast_{\rm qp}$. 
\endinsert
\vskip 28 truept

\centerline{\bf CONCLUSIONS}
\vskip 12 truept

In summary, we have performed a complete functional optimization 
of the density matrix of liquid $^4$He, based on a trial form of
generalized Jastrow type involving only two-point descriptors
of the dynamical correlations and the collective and single-particle 
thermal excitations.  The Euler-Lagrange equations of this approximate 
description have been solved numerically in the interesting 
temperature range $0\le T\le 5.5$~K.  The ansatz for the 
density matrix admits a simultaneous treatment of the Bose-Einstein 
condensed phase and the normal phase.  However, in the absence of
backflow effects, the predicted lambda transition temperature
of $T_\lambda\simeq 3.4$~K exceeds the experimental value of 2.18~K 
by some 1.2 K.  We argue that this shortcoming of the theory
stems from an inability of the chosen trial density matrix to accommodate
important variations of effective masses -- an inflexibility that 
results in high-lying collective excitations and quasihole and 
quasiparticle excitations that drive the theoretical transition 
temperature to unrealistically high values.  This view 
is strengthened by the results obtained when the optimization
is repeated in the normal phase with an adjustable effective mass 
for quasiparticles.  A renormalized mass $m_{\rm qp}^\ast=1.6m$
leads to a transition temperature nearly coincident with the 
experimental value. 

To pursue a systematic microscopic inclusion of backflow and effective
mass 
effects, the present treatment should be generalized to allow for
triplet pseudopotentials or three-point factors in the
temperature-dependent 
wave function (2) and the incoherence factor (3).  These extensions
will be required to achieve quantitative accuracy for the correlation 
energy and the dispersion relation of phonons and rotons.  Existing 
techniques [44,45] may be appropriated to carry through the associated 
analysis and computations, which should include effects of elementary 
diagrams.  In addition, a generalization of the ansatz (4) will be 
needed to take proper account of the quasiparticle and quasihole 
effective masses. The introduction of shadow coordinates [42,43] 
offers an alternate route to successful microscopic extension of 
the current realization of density-matrix theory.
\vskip 28 truept

\centerline{\bf ACKNOWLEDGMENTS}
\vskip 12 truept

The research reported herein has been supported by the U.S.\ National
Science Foundation under Grant No.\ PHY-9900713 and 
by the Deutsche Forschungsgemeinschaft, Graduiertenkolleg GRK549.

\vskip 28 truept
\centerline{{\bf APPENDIX}}
\vskip 12 truept
In this appendix, we collect the necessary formulae for
energies, occupation numbers, generalized structure functions,
and related quantities.

The ingredients $E_c$, $\epsilon_{\rm coll}$,  
$\epsilon_{\rm cc}$ of the energy functional defined by Eqs.~(5)--(8) 
have the explicit expressions
$$
\eqalignno{
E_{\rm c} &=  N {\rho\over 2}\int v^\ast(r)\, g(r) \, d\bfr \, & \cr
    &+{1\over2}\sum_\bfk \epsilon_{0}(k)\bigl[1 - \Theta(n_{c}(k))
   \bigl]\Gamma_{cc}(k)\left[1 + S^{(0)}_{cc}(k)+ 2\bcc S^{(0)}
(k)\right]\,,  &({\rm A1}) \cr}
$$  
where 
$$
 v^{\ast}(r) = v(r) - {\hbar^{2}\over 4 m}\Delta u(r)
$$
is the Feenberg effective potential [9,10], and
$$\eqalignno{
\epsilon_{\rm coll}(k) &= {1\over 2}\epsilon_{\rm o}(k)
\Bigg\{\Bigl[1+4\gamma(k) S(k)\Bigr]^{1/2}+1\Bigg\}S(k)^{-1}\,,&({\rm
A2})\cr
\epsilon_{\rm cc}(k) &=\epsilon_{\rm o}(k)\Biggl\{1 - \Bigl(1
  -X^{(0)}_{cc}(k)-\Gamma_{cc}(k)\Bigr)\Bigl[S_{cc}(k)-\tilde{S}_{cc}
   (k)\Bigr]\Biggr\} \,. &({\rm A3}) \cr}
$$
The thermal occupation numbers associated with collective,
quasihole, and quasiparticle excitations are given respectively by
$$\eqalignno{
n (k) &= {1\over2}\Bigl[\Bigl(1+4\gamma(k) S(k)\Bigr)^{1/2} - 1\Bigr]\,
        ,&({\rm A4})\cr
n_{cc}(k) &= \Gamma_{cc}(k)\Bigl[1+S^{(0)}_{cc}(k)\Bigr]\,,&({\rm A5})\cr
n_{c}(k) &= \Gamma_{cc}(k)\Bigl[1 + S^{(0)}_{cc}(k)+2\bcc 
S^{(0)}(k)\Bigr]_+ \, , &({\rm A6})\cr }
$$
where $[F]_+$ delivers $F$ itself if $F \geq 0$ and zero otherwise.
The structure functions $S(k)$ and $S_{cc}(k)$ and their 
various components, as well as the corresponding Fourier inverses
$g(r)=1+G(r)$ and $G_{cc}(r)$, etc., are defined in Ref.~[25]. 
They may be evaluated by means of the hypernetted-chain equations 
contained therein.

The functions $D(k)$ and $D_{cc}(k)$ appearing in the Euler-Lagrange 
equations (15) and (16) are constructed as [28]
$$
\eqalignno{
D(k) &= -{2\over N}\sum_{\bfk'} {\bf T}(k')\cdot {\delta {\bf S}_{D}
        (k')\over\delta u (\bfk)}-\lambda\bcc {\partial\over\partial \bcc}
   S(k)\, ,&({\rm A7})\cr}
$$
$$
\eqalignno{
D_{cc}(k) &= -{1\over N}\sum_{\bfk'} {\bf T}(k')\cdot {\delta 
  {\bf S}_{D}(k')\over\delta\Gamma_{cc}(\bfk)} + 
   \Bigl[{1\over2}\omega_{cc}(k)+\lambda\Bigr]\Bigl[1+S^{(0)}_{cc}(k)
   \Bigr]\, & \cr
  & + {1\over2}\omega_{c}(k)\Biggl\{1+S^{(2)}_{cc}(k)+2\bcc \Bigl[S_{dd}
    (k)+2S_{dc}(k)\Bigr]\Biggr\}\Theta\bigl(n_{c}(k)\bigr)\, & \cr
  & - \epsilon_{0}(k)\Bigl[1+ \tilde{S}_{cc}(k)\Bigr]\, & \cr
  & - \lambda\bcc\Biggl\{{\partial\over\partial\bcc}S^{(0)}_{cc}(k)+ \bcc
    {\partial\over\partial\bcc}\Bigl[S_{dd}(k)+2S_{dc}(k)+S^{(1)}_{cc}
    (k)\Bigr]\Biggr\}  \, . &({\rm A8}) \cr}
$$  
The quantity ${\bf S}_{D}(k) = \bigl( S_{dd}(k),S_{de}(k),S_{ee}(k), 
S_{dc}(k),S_{ec}(k),S_{cc}(k) \bigr)$
is defined by the indicated sextuple of components.
In Eqs.~(A7) and (A8) these components form a scalar product with 
the six components $T^{(i)}$ of the quantity 
${\bf T}(k), i=1,2,3,...,6$.  The component $T^{(1)}$ is made up
of two terms,
$$
\eqalignno{
T^{(1)}(k) &=  T^{(1)}_{1}(k)+  T^{(1)}_{2}(k)\, ,& ({\rm A9})\cr}
$$
the first being the Fourier transform
$$
\eqalignno{
T^{(1)}_{1}(k) &= \rho\int T^{(1)}_{1}({\bf r}) 
e^{i{\bf k}\cdot{\bf r}} d{\bf r}
\, & ({\rm A10}) \cr}
$$
of
$$\eqalignno{
T^{(1)}_{1}(\rij)&= {\hbar^2\over 8m}\Biggl\{\Bigl[{\bf\nabla}_{1}\Bigl(
\tilde{N}^{(0)}_{cc}(\rij)+ \Gamma_{cc}(\rij)\Bigr)\Bigr]^2 \Bigr.\, & \cr
   &+{\bf\nabla}_{1}\tilde{N}_{cd}(\rij)\Biggl(4\bcc^2 N_{dc}(\rij)
    {\bf\nabla}_{1}\tilde{N}_{cd}(\rij)\Bigr.\, & \cr
   & +2\bcc\Bigl(1+N_{dc}(\rij)\Bigr)\Bigl[ 2\bcc N_{dc}(\rij)
    {\bf\nabla}_{1}\tilde{N}_{cd}(\rij)\Bigr.\, &\cr
   &\Bigl.\Bigl. + {\bf\nabla}_{1}
   \Bigl(\tilde{N}^{(2)}_{cc}(\rij)+\gamcc(\rij)\Bigr)\Bigr]\Biggr)\, &
\cr
   &+{\bf\nabla}_{1}\Bigl(\tilde{N}^{(2)}_{cc}(\rij)+\gamcc(\rij)\Bigr)
   \Bigl[\Bigl(1+N_{dc}(\rij)\Bigr){\bf\nabla}_{1}\tilde{N}_{cd}(\rij) 
   \Bigr.\, &\cr
   & \Bigl.\Bigl.+ {\bf\nabla}_{1}
   \Bigl(\tilde{N}^{(2)}_{cc}(\rij)+\gamcc(\rij)\Bigr) \Bigr]\Biggr\}\,,
    & ({\rm A11})\cr}
$$
and the second being given by
$$\eqalignno{
T^{(1)}_{2}(k)&= {1\over2}\bcc\epsilon_{0}(k)\Biggl[\tilde{X}_{cd}(k)+
\tilde{X}_{ce}(k)+ \tilde{X}^{(2)}_{cc}(k)+\Gamma_{cc}(k)\Biggr]
   \, &\cr
  &\Biggl\{\tilde{S}^{(2)}_{cc}(k)\Bigl[1- X_{dd}(k)- X_{de}(k)-2\bcc 
   X_{dc}(k)\Bigr]\Bigr.\, & \cr
 & \Bigl.-\Bigl[2\bcc\tilde{S}_{cd}(k)+\tilde{S}^{(2)}_{cc}(k)\Bigr]
\Bigl[X_{dc}
 (k)+X_{ec}(k)+ X^{(2)}_{cc}(k)+\Gamma_{cc}(k)\Bigr]\Biggr\} 
\,& \cr
 & + \bcc \Gamma_{cc}(k)\Bigl[\omega_{cc}(k)-\omega_{c}(k)\Theta(n_{c}
  (k))\Bigr]\,& \cr
  &-{1\over2}\bcc\epsilon_{0}(k)\Bigl[\tilde{X}^{(0)}_{cc}(k)+\Gamma_{cc}
  (k)\Bigr]^2 \,.& ({\rm A12})\cr}
$$
For $T^{(2)}$ and $T^{(3)}$ we have
$$
\eqalignno{
T^{(2)}(k) &= {1\over2}\bcc\epsilon_{0}(k)\Biggl\{\tilde{S}^{(2)}_{cc}
(k)\tilde{X}_{cd}(k)\Bigl[1- X_{dd}(k)- X_{de}(k)-2\bcc X_{dc}
(k)\Bigr]\Bigr.\, & \cr
 &-\tilde{X}_{cd}(k)\Bigl[2\bcc\tilde{S}_{cd}(k)+
\tilde{S}^{(2)}_{cc}
(k)\Bigr]\, \Bigl[X_{dc}(k)+X_{ec}(k)+ X^{(2)}_{cc}(k)+\Gamma_{cc}
 (k)\Bigr]\, & \cr
 &-\Bigl[\tilde{S}^{(2)}_{cc}(k)X_{dd}(k)+ \Bigl(2\bcc
\tilde{S}_{cd}(k)+\tilde{S}^{(2)}_{cc}(k)\Bigr)X_{dc}(k)\Bigr]
 \,&\cr
 &\times\Bigl.\Bigl(\tilde{X}_{cd}(k)+\tilde{X}_{ce}(k)+
 \tilde{X}^{(2)}_{cc}(k)+ \gamcc(k) \Bigr)\Biggr\}  & ({\rm A13})\cr}
$$
and
$$
T^{(3)}(k) = -{1\over2}\bcc\epsilon_{0}(k)\tilde{X}_{cd}(k)\Biggl\{
  \tilde{S}^{(2)}_{cc}(k)X_{dd}(k)
 +\Bigl(2\bcc\tilde{S}_{cd}(k)+
  \tilde{S}^{(2)}_{cc}(k)\Bigr)X_{dc}(k) \Biggr\} \, . \eqno({\rm A14})
$$
The component $T^{(4)}$ again consists of two parts,
$$
\eqalignno{
T^{(4)}(k) &=  T^{(4)}_{1}(k)+  T^{(4)}_{2}(k)\, ,& ({\rm A15})\cr}
$$
the first being the Fourier transform
$$\eqalignno{
T^{(4)}_{1}(k)&= \rho\int T^{(4)}_{1}(r) 
e^{i{\bf k} \cdot {\bf r}} d{\bf r} & ({\rm A16}) \cr}
$$
of
$$
\eqalignno{
T^{(4)}_{1}(\rij)&={\hbar^2\over 8m}\left(2\bcc
 {\bf\nabla}_{1} \tilde{N}_{cd}(\rij)\right) \,&\cr
&\times\left\{2\bcc 
\left(\,1+N_{cd}(\rij)\right)\,
{\bf\nabla}_{1}\tilde{N}_{cd}(\rij) +
\tilde{N}^{(2)}_{cc}(\rij)+\gamcc(\rij)
\right\} \, & ({\rm A17})\cr}
$$
and the second having the expression
$$
\eqalignno{
T^{(4)}_{2}(k)&= {1\over2}\bcc\epsilon_{0}(k)
\Biggl\{\biggl[\tilde{S}^{(2)}_{cc}(k)\Bigr.\Bigr.
\Bigl(1-X_{dd}(k)- 
X_{de}(k)-2\bcc X_{dc}(k)\Bigr)\, & \cr
&-\Bigl.\Bigl(2\bcc\tilde{S}_{cd}(k)+\tilde{S}^{(2)}_{cc}(k)\Bigr)
\Bigl(X_{dc}(k)+X_{ec}(k)+ X^{(2)}_{cc}(k)+\Gamma_{cc}(k)
    \Bigr)\biggr]\, & \cr
&\times \Bigl(2\bcc\tilde{X}_{cd}(k)+\tilde{X}^{(2)}_{cc}(k)+
  \gamcc(k)\Bigr)\, & \cr
&+\biggl[\Bigl(2\bcc\tilde{S}_{cd}(k)+\tilde{S}^{(2)}_{cc}(k)\Bigr)
 \Bigl(1-2\bcc X_{dc}(k)-X^{(2)}_{cc}(k)-\gamcc(k)\Bigr)\Bigr.\, & \cr
&\Bigl.\Bigl. -2\bcc\tilde{S}^{(2)}_{cc}(k)
\Bigl(X^{}_{dd}(k)+X_{dc}(k)\Bigr)
\biggr]\Bigl(\tilde{X}_{cd}(k)+\tilde{X}_{ce}(k)+\tilde{X}^{(2)}_{cc}(k)
+\gamcc(k)\Bigr) \Biggr\} & \cr
&+2\bcc\biggl[\Gamma_{cc}(k)\Bigl( \omega_{cc}(k)-\omega_{c}(k)\Theta
 (n_{c}(k))\Bigr)\Bigr. \cr
& \Bigl. -{1\over2}\epsilon_{0}(k)\Bigl(\tilde{X}^{(0)}_{cc}
 (k)+\Gamma_{cc}(k)\Bigr)^2\biggr] \, .&\, ({\rm A18})\cr}
$$
The fifth and sixth components read
$$
\eqalignno{
T^{(5)}(k) &= {1\over2}\bcc\epsilon_{0}(k)\Biggl\{\tilde{X}_{cd}
 (k)\biggl[\Bigl(2\bcc\tilde{S}_{cd}(k)+\tilde{S}^{(2)}_{cc}(k)
 \Bigr)\Bigr.\Bigr.\, &\cr
 &\times\Bigl(1-2\bcc X_{dc}(k)-X^{(2)}_{cc}(k)-\gamcc(k)\Bigr)
 \, &\cr
&\Bigl. -2\bcc\tilde{S}^{(2)}_{cc}(k)\Bigl(X_{dd}(k)+X_{dc}(k)
 \Bigr)\biggr]\, &\cr
&-\biggl(\tilde{S}^{(2)}_{cc}(k)X_{dd}(k)+\Bigl(2\bcc
\tilde{S}_{cd} (k)+\tilde{S}^{(2)}_{cc}(k)\Bigr)X_{dc}(k) \biggr)
 \, &\cr
 &\times\Bigl.\Bigl(2\bcc\tilde{X}_{cd}(k)+\tilde{X}^{(2)}_{cc}(k)
+\gamcc(k)\Bigr)\Biggr\}\, ,& ({\rm A19})\cr}
$$
$$
\eqalignno{
T^{(6)}(k) &= {1\over4}\epsilon_{0}(k)\Biggl\{\Bigl(2\bcc
\tilde{S}_{cd}(k)+\tilde{S}^{(2)}_{cc}(k)\Bigr)\Bigr.\,&\cr
&\times \Bigl(1-2\bcc X_{dc}(k)-X^{(2)}_{cc}(k)-\gamcc(k)\Bigr)
 \, &\cr
&\Bigl. -2\bcc\tilde{S}^{(2)}_{cc}(k)\Bigl(X_{dd}(k)+X_{dc}(k)
\Bigr)\Biggr\}\Bigl(2\bcc\tilde{X}_{cd}(k)+\tilde{X}^{(2)}_{cc}
 (k)+\gamcc(k)\Bigr)\, &\cr
&+{1\over2}\biggl[\Gamma_{cc}(k)\Bigl(\omega_{cc}(k)-\omega_{c}(k)
\Theta\bigl(n_{c}(k)\bigr)\Bigr)\, &\cr
&-{1\over2}\epsilon_{0}(k)\Bigl(\tilde{X}^{(0)}_{cc}
 (k)+\Gamma_{cc}(k)\Bigr)^2\biggr] \, .& ({\rm A20})\cr}
$$
For the normal phase ($\bcc\equiv 0$), these expressions simplify
considerably, since many terms vanish [28].

The generalized structure functions $\tbul{S}(k)$ and $\tbul{S}_{cc}(k)$
entering the Euler-Lagrange equations (15) and (16) may be
explicated using the general prescription [28]
$$\eqalignno{
\tbul{S}_{\alpha\beta}(k)& ={1\over N}\sum_{\bfk'}\Bigl(v^{\ast}(k')+ 
 v^{\ast}_{coll}(k')\Bigr){\delta S_{\alpha\beta}(k)\over
 \delta  u(\bf{k}')} \, & \cr
 & +{1\over
N}\sum_{\bfk'}\Biggl\{\gamcc(k')\Bigl({1\over2}\epsilon_{0}(k')-
 \omega_{cc}(k')-\lambda\Bigr)\Bigr.\, & \cr
 & \Bigl.+
{1\over2}\epsilon_{0}(k')\Bigl(\tilde{X}^{(0)}_{cc}(k')+\gamcc(k')
   \Bigr)^{2}\Biggr\}{\delta S_{\alpha\beta}(k)\over\delta\gamcc(\bfk')} 
    \,, & ({\rm A21})\cr}
$$
where $\alpha\beta=dd,$ $de$, $ee$, $ec$, $dc$, or $cc$ (see Ref.~[25]).
The dotted function $ \tbul{S}_{\alpha\beta}$ may be analyzed and
evaluated within the extended Fermi hypernetted-chain formalism [28].

\vskip 28 truept
\centerline{\bf REFERENCES}
\vskip 12 truept

\item{[1]}
P. Kapitza, {\it Nature} {\bf 141}, 74 (1938).
\smallskip

\item{[2]}
J. F. Allen and A. D. Misener, {\it Nature} {\bf 141}, 75 (1938).
\smallskip

\item{[3]}
F. London, {\it Nature} {\bf 141}, 643 (1938).
\smallskip

\item{[4]}
F. London, {\it Phys. Rev.} {\bf 54}, 947 (1938).
\smallskip

\item{[5]}
L. Tisza, {\it Nature} {\bf 141}, 913 (1938).
\smallskip

\item{[6]}
L. Tisza,
{\it Phys. Rev.} {\bf 72}, 838 (1947).
\smallskip

\item{[7]}
L. D. Landau, {\it Zh.~Eksp.~Teor.~Fiz.} {\bf 11}, 592 (1941); 
{\it J.~Phys.~ USSR} {\bf 5}, 71 (1941).

\item{[8]} R. P. Feynman, {\it Phys. Rev.} {\bf 91}, 1291 (1953).
\smallskip

\item{[9]}
C. E. Campbell and E. Feenberg, {\it Phys. Rev.} {\bf 188}, 396 (1969).
\smallskip

\item{[10]}
E. Feenberg, {\it Theory of Quantum Fluids} (Academic Press, New York,
1969).
\smallskip

\item{[11]}
E. Feenberg, {\it Ann. Phys. (NY)} {\bf 84}, 128 (1974).
\smallskip

\item{[12]}
R. M. Ziff, G. E. Uhlenbeck, and M. Kac, {\it Phys. Rep.} {\bf 32}, 169 
(1977).
\smallskip

\item{[13]}
C. E. Campbell, in {\it Recent Progress in Many-Body Theories}, Vol.~4
edited by E. Schachinger, H. Mitter, and H. Sormann (Plenum, New York,
1995), p. 29.

\item{[14]}
D. M. Ceperley and E. L. Pollock, {\it Phys. Rev. Lett.} {\bf 56}, 351
(1986).
\smallskip

\item{[15]}
D. M. Ceperley, {\it Rev. Mod. Phys.} {\bf 67}, 279 (1995).
\smallskip

\item{[16]}
S. Fantoni and S. Rosati, {\it Nuovo Cimento} {\bf 25A}, 593 (1975).
\smallskip

\item{[17]}
E. Krotscheck and M. L. Ristig, {\it Nucl. Phys.} {\bf A242}, 389 (1975).
\smallskip

\item{[18]}
K. Hiroike, {\it Prog. Theor. Phys.} {\bf 24}, 317 (1960).
\smallskip

\item{[19]}
M. L. Ristig, T. Lindenau, M. Serhan, and J. W. Clark,
{\it J. Low Temp.~Phys.} {\bf 114}, 317 (1999).
\smallskip

\item{[20]}
D. Pines, {\it Recent Progress in Many-Body Theories}, Springer Lecture
Notes in Physics {\bf 142}, edited by J. G. Zabolitzky, M. de Llano, 
M. Fortes, and J. W. Clark (Springer-Verlag, Berlin, 1981), p. 202.
\smallskip

\item{[21]}
J. Bardeen, {\it Physics Today} {\bf 16}, 21 (January 1963); quoted in
{\it Physics Today} {\bf 51}, 66 (May 1998).
 \smallskip

\item{[22]}
C. E. Campbell, K. E. K\"urten, M. L. Ristig, and G. Senger, 
{\it Phys. Rev. B} {\bf 30}, 3728 (1984).
\smallskip

\item{[23]}
G. Senger, M. L. Ristig, K. E. K\"urten, and C. E. Campbell, 
{\it Phys.~Rev.~B}~{\bf 33}, 7562 (1986).
\smallskip

\item{[24]} 
G. Senger, M. L. Ristig, C. E. Campbell, and 
J. W. Clark, {\it Ann.~Phys.~(N.~Y.)} {\bf 218}, 116 (1992).
\smallskip

\item{[25]} 
M. L. Ristig, G. Senger, M. Serhan, and J. W. Clark,
{\it Ann. Phys.~(N.~Y.)} {\bf 243}, 247 (1995).
\smallskip

\item{[26]} 
J. W. Clark, M. L. Ristig, T. Lindenau, and M. Serhan, in 
{\it Condensed Matter Theories}, Vol.\ 12, edited by J. W. Clark and 
P. V. Panat (Nova Science Publishers, Commack, NY, 1997), 
p.\ 55.
\smallskip

\item{[27]}
M. L. Ristig, T. Lindenau, M. Serhan, and J. W. Clark, in {\it 
Condensed Matter Theories}, Vol.\ 13,  edited by J. da Providencia 
and F.~B.~Malik (Nova Science Publishers, Commack, NY, 1998), 
p.\ 119.
\smallskip

\item{[28]} 
T. Lindenau, Doctoral thesis, Universit\"at zu K\"oln  
(Shaker Verlag, Aachen, 1999).
\smallskip

\item{[29]}
R. P. Feynman, {\it Phys. Rev.} {\bf 94}, 262 (1954).
\smallskip

\item{[30]}
E. Krotscheck, in {\it 150 Years of Quantum Many-Body Theory}, 
edited by R. F. Bishop, K. A. Gernoth, and N. R. Walet (World
Scientific, Singapore, 2001).
\smallskip

\item{[31]}
W. Kohn and L. J. Sham, Phys. Rev. {\bf 140}, A1133 (1965).

\item{[32]}
J. W. Clark, in {\it Progress in Particle and Nuclear Physics}, Vol.\ 2, 
edited by D. Wilkinson (Pergamon, Oxford, 1979), p.\ 89.
\smallskip

\item{[33]}
R. A. Aziz, V. P. S. Nain, J. S. Carley, W. L. Taylor, and G. T.
McConville, {\it J. Chem. Phys.} {\bf 70}, 4330 (1979).
\smallskip

\item{[34]}
R. P. Feynman, {\it Statistical Mechanics} (Benjamin, Reading, 1972),
p.~318.
\smallskip

\item{[35]}
R. K. Crawford, in {\it Rare Gas Solids}, Vol.\ 2, edited by M. L. Klein 
and J. A. Venables (Academic Press, New York, 1976).
\smallskip

\item{[36]}
C. E. Campbell, {\it Phys. Lett. A} {\bf 44}, 471 (1973).
\smallskip

\item{[37]}
E. Krotscheck, in {\it Lecture Notes in Physics}, Vol.\ 510, 
edited by J. Navarro and A. Polls (Springer, Heidelberg, 1998).
\smallskip

\item{[38]}
K. Schmidt, M. H. Kalos, Michael A. Lee, and G. V. Chester, 
{\it Phys. Rev. Lett.} {\bf 45}, 573 (1980).
\smallskip

\item{[39]} 
R. Pantf\"order, T. Lindenau, and M. L. Ristig, {\it J. Low 
Temp.~Phys.}~{\bf 108}, 245 (1997).
\smallskip

\item{[40]}
T. Lindenau, M. L. Ristig, and J. W. Clark, in {\it Condensed Matter 
Theories}, Vol.\ 14, edited by D.~Ernst, I. Perakis, and S. Umar 
(Nova Science Publishers, Huntington, New York, 1999), p.131.  
\smallskip

\item{[41]}
M. H. Kalos, M. A. Lee, P. A. Whitlock, and G. V. Chester, 
{\it Phys. Rev.~B} {\bf 24}, 115 (1981).
\smallskip

\item{[42]}
S. Vitiello, K. Runge, and M. H. Kalos, {\it Phys. Rev. Lett.} {\bf 60},
1970 (1988).
\smallskip

\item{[43]}
L. Reatto and G. L. Masserini, {\it Phys. Rev. B} {\bf 38}, 4516 (1988).
\smallskip

\item{[44]}
R. P. Feynman and M. Cohen, {\it Phys. Rev.} {\bf 102}, 1189 (1956).
\smallskip

\item{[45]}
B. E. Clements, E. Krotscheck, J. A. Smith, and C. E. Campbell, {\it Phys. 
Rev. B} {\bf 47}, 5239 (1993). 
\end